\documentclass[aps,prl,twocolumn,preprintnumbers,amsmath,amssymb,floats,citeautoscript]{revtex4-1}
\usepackage{graphicx}

\begin{document}
\preprint{Science {\bf 331}, 439--442 (2011).}
 
\title{Rotational Symmetry Breaking in the Hidden-Order Phase of URu$_2$Si$_2$}

\author{R.~Okazaki$^{1,*}$, T.~Shibauchi$^{1,\dag}$, H.\,J.~Shi$^{1}$, Y.~Haga$^{2}$, T.~D.~Matsuda$^{2}$, E.~Yamamoto$^{2}$, Y.~Onuki$^{2,3}$, H.~Ikeda$^{1}$,  and Y.~Matsuda$^{1}$}
\affiliation{
$^1$Department of Physics, Kyoto University, Kyoto 606-8502, Japan\\
$^2$Advanced Science Research Center, Japan Atomic Energy Agency, Tokai 319-1195, Japan\\
$^3$Graduate School of Science, Osaka University, Toyonaka, Osaka 560-0043, Japan\\
$^*$Present address: Department of Physics, Nagoya University, Nagoya 464-8602, Japan\\
$^\dag$To whom correspondence should be addressed. E-mail: {\sf\small shibauchi@scphys.kyoto-u.ac.jp}
}

\date{published January 28, 2011}

\begin{abstract}
{\bf
A second-order phase transition is characterized by spontaneous symmetry breaking. The nature of the broken symmetry in the so-called ``hidden order'' phase transition in the heavy fermion compound URu$_2$Si$_2$, at transition temperature $T_h=17.5$\,K, has posed a long-standing mystery. We report the emergence of an in-plane anisotropy of the magnetic susceptibility below $T_h$, which breaks the four-fold rotational symmetry of the tetragonal URu$_2$Si$_2$. Two-fold oscillations in the magnetic torque under in-plane field rotation were sensitively detected in small pure crystals. Our findings suggest that the hidden-order phase is an electronic ``nematic'' phase, a translationally invariant metallic phase with spontaneous breaking of rotational symmetry.
}
\end{abstract}



\maketitle

A second-order phase transition generally causes a change in symmetry, such as rotational, gauge, or time-reversal symmetry. An order parameter can then be introduced to describe the low-temperature ordered phase with a reduced symmetry. 
The heavy-fermion compound URu$_2$Si$_2$ undergoes a second-order phase transition at $T_h=17.5$\,K, which is accompanied by large anomalies in thermodynamic and transport properties \cite{Pal85,Map86,Ram92}. Because the nature of the associated order parameter has not been elucidated, the low-temperature phase is referred to as the hidden-order phase. It is characterized by several remarkable features. No structural phase transition is observed at $T_h$. A tiny magnetic moment appears ($M_0 \sim 0.03\mu_B$, where $\mu_B$is the Bohr magneton) below $T_h$ \cite{Bro87}, but it is far too small to explain the large entropy released during the transition and seems to have an extrinsic origin \cite{Mat01,Tak07,Ami07}. An electronic excitation gap is formed on a large portion of the Fermi surface and most of the carriers ($\sim 90\%$) disappear \cite{Sch87,Beh05,Kas07}; a gap is also formed in the incommensurate magnetic excitation spectrum \cite{Wie07}. 

The nature of the hidden order cannot be determined without understanding which symmetry is being broken. Several microscopic models, including multipole ordering \cite{San94,Kis05,Hau09,Cri09,Har10}, spin-density wave formation \cite{Ike98,Min05,Elg09}, orbital currents \cite{Cha02}, and helicity order \cite{Var06}, have been proposed. However, despite intensive experimental and theoretical studies, this remains an open question.  

The measurement of the magnetic torque {\boldmath $\tau$} = $\mu_0${\boldmath $M$}$V \times${\boldmath $H$} has a high sensitivity for detecting magnetic anisotropy. Here $V$ is the sample volume, {\boldmath $M$} is the induced magnetization, {\boldmath $H$} is the magnetic field, and $\mu_0$ is the permeability of vacuum. In particular, torque measurements performed for a range of directions of {\boldmath $H$} spanning the tetragonal $ab$ plane in URu$_2$Si$_2$ provide a stringent test of whether the hidden-order parameter breaks the crystal four-fold symmetry. In such a geometry, $\tau$ is a periodic function of double the azimuthal angle $\phi$ measured from the $a$ axis:
\begin{equation}
\tau_{2\phi} = \frac{1}{2}\mu_0H^2V\left[(\chi_{aa}^{}-\chi_{bb}^{})\sin2\phi - 2\chi_{ab}\cos2\phi\right],
\end{equation}
where the susceptibility tensor $\chi_{ij}$ is given by $M_i=\displaystyle\sum_j \chi_{ij} H_j$ (Fig.\:1). In a system holding tetragonal symmetry, $\tau_{2\phi}$ should be zero because $\chi_{aa}=\chi_{bb}$ and $\chi_{ab}=0$. Finite values of $\tau_{2\phi}$ may appear if a new electronic or magnetic state emerges that breaks the tetragonal symmetry. In such a case, rotational symmetry breaking is revealed by $\chi_{aa}\ne\chi_{bb}$ or $\chi_{ab}\ne0$ depending on the orthorhombicity direction. 

\begin{figure}[b]
\includegraphics[width=1.05\linewidth]{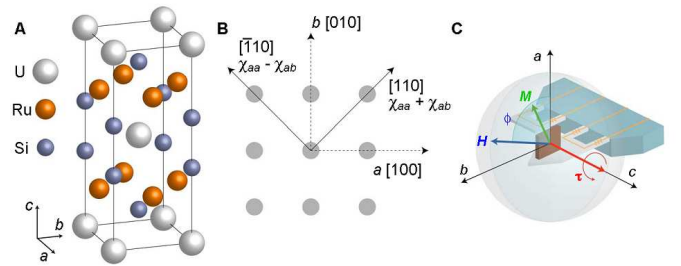}
\caption{({\bf A}) The tetragonal crystal structure of URu$_2$Si$_2$. ({\bf B}) U atom arrangement in the $ab$ plane. The relevant axes and the susceptibility components for the $\chi_{aa}=\chi_{bb}$ case are also shown. ({\bf C}) Schematics of in-plane $\phi$-scan measurements. The magnetic field {\boldmath $H$} was applied in the $ab$ plane with high alignment precision (within 0.02$^\circ$).
}
\end{figure}

Figure\:2A depicts the torque measured in field {\boldmath $H$}, the orientation of which is varied within a plane that includes the $c$ axis (Fig.\:2A, inset). Both below and above $T_h$, the curves are perfectly sinusoidal and can be fitted with $\tau(T,H,\theta)=A_{2\theta}(T,H)\sin 2\theta$, where $A$ is the amplitude, $\theta$ is the polar angle, and $T$ is absolute temperature. In our carefully selected crystals \cite{SOM}, no hysteresis is observed, indicating no detectable ferromagnetic impurities (the hysteresis component is less than 0.01\% of the total torque).  In this geometry, the difference $\Delta\chi_{ca}$ between the $c$-axis and in-plane susceptibilities yields a two-fold oscillation term $\tau_{2\theta}(\theta,T,H)$ with respect to $\theta$ rotation,
\begin{equation}
\tau_{2\theta} = \frac{1}{2}\mu_0H^2V\Delta\chi_{ca}\sin2\theta.
\end{equation}

The $H$-linear dependence of $A_{2\theta}(H,T)/H$ with a negligible $y$-intercept (Fig.\:2B) indicates a field-independent magnetic susceptibility--- that is, a purely paramagnetic response.  This also reinforces the absence of ferromagnetic impurities.  To check the consistency of the data, we also measured the susceptibility of a different single crystal of URu$_2$Si$_2$ in the same batch by means of a superconducting quantum interference device (SQUID) magnetometer (Fig.\:2C). The amplitudes of $\Delta\chi_{ca}$ obtained in the two types of measurements coincide in both the hidden-order and paramagnetic states.  

\begin{figure}[t]
\includegraphics[width=1.05\linewidth]{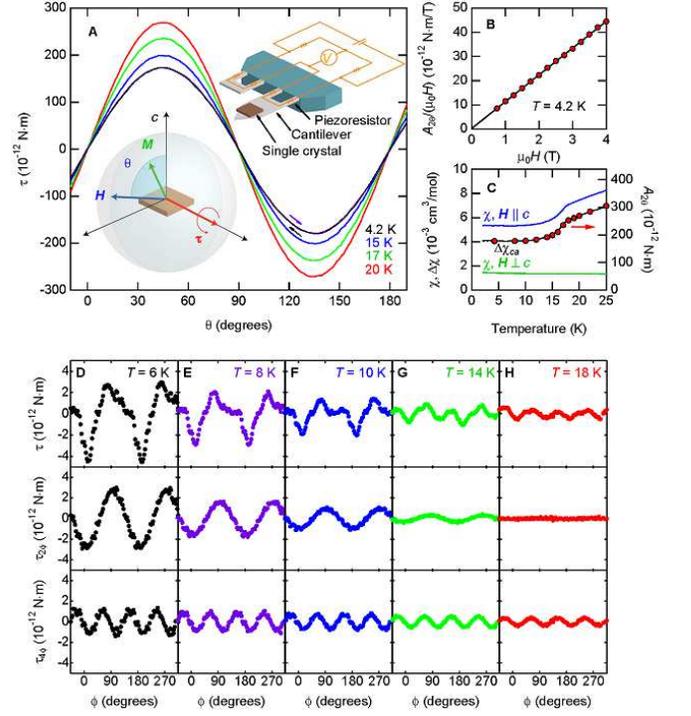}
\caption{
Magnetic torque curves measured at $\left|\mu_0\mbox{\boldmath $H$}\right| = 4$\,T. ({\bf A}) Torque $\tau(\theta)$ as a function of the polar angle $\theta$ (lower inset) measured at several temperatures. Torque curves measured by rotating {\boldmath $H$} in clockwise (dotted line) and anticlockwise (soli line) directions coincide. Upper inset illustrates the experimental configuration for $\tau(\theta)$ measurements. ({\bf B}) Magnetic field dependence of the amplitude of the two-fold oscillation of the torque divided by the field, $A_{2\theta}/(\mu_{0}H)$, at $T$ = 4.2 K. ({\bf C}) Left axis: Temperature dependence of the susceptibility $\chi$ for {\boldmath $H$} $\perp c$ (green) and {\boldmath $H$} $\parallel c$ (blue). Their difference $\Delta\chi_{ca}$ is shown in black. Right axis: Temperature dependence of $A_{2\theta}$ (red circles). ({\bf D} to {\bf H}) Upper panels show raw torque curves $\tau(\phi)$ as a function of the azimuthal angle $\phi$ at several temperatures. Middle and lower panels show the two-fold $\cos2\phi$ and four-fold $\sin4\phi$ components of the torque curves, respectively, obtained from Fourier analysis of the raw torque curves. 
}
\end{figure}

Having established the evidence of the purely paramagnetic response in our single crystals,  we now consider the in-plane torque measured in {\boldmath $H$} rotating within the $ab$ plane. To exclude two-fold oscillations appearing as a result of misalignment, {\boldmath $H$} is precisely applied in the $ab$ plane within an error of less than 0.02$^\circ$ by controlling two superconducting magnets and a rotating stage \cite{SOM} (Fig.\:1C). Shown in Fig.\:2, D to H, is the temperature evolution of the torque $\tau(\phi)$ at 4\,T in the hidden-order phase.  All torque curves are perfectly reversible with respect to the field rotation direction; $\tau(\phi)$ can be decomposed as $\tau=\tau_{2\phi}+\tau_{4\phi}+\tau_{6\phi}+\cdots$, where $\tau_{2n\phi}=A_{2n\phi}\sin 2n(\phi-\phi_0)$ is a term with $2n$-fold symmetry with $n=1,2,\cdots$.  In the middle and lower panels of Fig.\:2, D to H, the two-fold and four-fold components obtained from the Fourier analysis are displayed.

Our results show that the two-fold oscillation is distinctly observed in the hidden-order state, whereas it is absent in the paramagnetic phase at $T = 18$\,K slightly above $T_h$ (middle panels of Fig.\:2, D to H). Indeed, the torque curves shown in the upper panels become asymmetric with respect to $90^\circ$ rotations below $T_h$.  We note that the observed four-fold oscillations $\tau_{4\phi}$ (and higher-order terms) arise primarily from the nonlinear susceptibilities \cite{Mor84}. The presence of the two-fold oscillation, which follows the functional form $\tau_{2\phi}=A_{2\phi}\cos 2\phi$, clearly demonstrates that $\chi_{ab}\ne0$, whereas $\chi_{aa}=\chi_{bb}$. Figure\:3A depicts the temperature dependence of $|A_{2\phi}|/V$, which indicates sizable in-plane anisotropy $2\chi_{ab}/\chi_{aa}$ at low temperatures. The two-fold amplitude becomes nonzero precisely at $T_h=17.5$\,K (Fig.\:3A, inset).  This result, together with the absence of hysteresis in the torque curves, rules out the possibility that very tiny ferromagnetic impurities are the origin of the two-fold symmetry \cite{SOM}. Moreover, $|A_{2\phi}|$ grows rapidly with decreasing temperature in a manner quite different from $\Delta\chi_{ca}(T)$ (Fig.\:2C). This indicates that a misalignment of {\boldmath $H$} from the $ab$ plane is unlikely to be the origin of the two-fold symmetry. On the basis of this reasoning, we conclude that the amplitude of two-fold oscillations is a manifestation of intrinsic in-plane anisotropy of the susceptibility (Fig.\:1B): 
\begin{equation}
\chi[110]=\chi_{aa}+\chi_{ab} \ne \chi[\bar{1}10]=\chi_{aa}-\chi_{ab}.  
\end{equation}

The in-plane anisotropy that sets in precisely at $T_h$ indicates that the rotational symmetry is broken in the hidden-order phase. The temperature dependence of $|A_{2\phi}| \propto \chi_{ab}$ near $T_h$ is fitted with $\chi_{ab} \propto c(T_{h}-T)+c'(T_{h}-T)^{2}$ much better than with a purely quadratic fit (Fig.\:3A, inset). The leading $T$-linear term near $T_h$ (Fig.\:3B) is naturally expected from the standard Landau theory of second-order transition when $\chi_{ab}$ is described by the squared term of an order parameter $\eta \propto (T_{h}-T)^{1/2}$. Thus, the observed in-plane anisotropy $2\chi_{ab}/\chi_{aa}$ implies that the hidden-order parameter $\eta$ breaks the four-fold tetragonal symmetry.

\begin{figure}[t]
\includegraphics[width=1.1\linewidth]{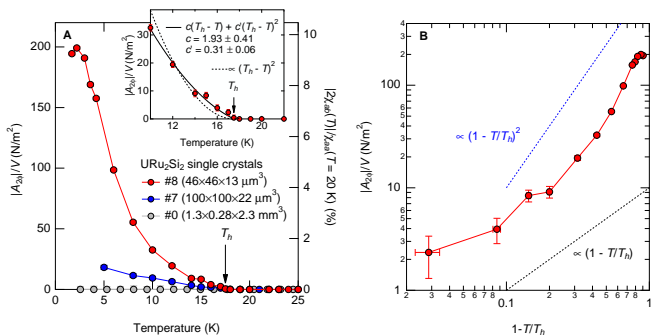}
\caption{
({\bf A}) Temperature dependence of two-fold oscillation amplitude divided by the sample volume $|A_{2\phi}|/V$ measured for sample 7 (blue circles) and sample 8 (red circles). The normalized in-plane susceptibility anisotropy $2\chi_{ab}/\chi_{aa}=(\chi[110]-\chi[\bar{1}10])/\chi[100]$ is evaluated in the right axis. SQUID data for the large crystal, sample 0, are also shown (gray circles) in the same scale. Inset: the data for sample 8 near $T_{h}=17.5$\,K are fitted to $c(T_{h}-T)+c'(T_{h}-T)^{2}$ (solid line) and the quadratic dependence $\propto(T_{h}-T)^{2}$ (dashed line).  ({\bf B}) Temperature variation of $|A_{2\phi}|/V$ for sample 8 on a log-log scale. The black and blue dotted lines represent linear and quadratic temperature dependence, respectively.
}
\end{figure}

Why has such an in-plane magnetic anisotropy not been reported before? To address this, we measured several samples with different sizes. In millimeter-sized crystals, we observed no difference between $\chi[110](T)$ and $\chi[\bar{1}10](T)$, but in samples with a smaller volume $V$, a nonzero $2\chi_{ab}\propto |A_{2\phi}|/V$ appeared (Fig.\:3A). This may imply that the hidden-order phase forms domains with different preferred directions in the $ab$ plane, which may be a natural consequence of the tetragonal crystal structure (with possible small undetected orthorhombicity in the domain). A domain size on the order of tens of microns would explain both our results and the difficulties in observing this effect. The fact that the torque curves remain unchanged for field-cooling conditions at different field angles \cite{SOM} implies that the formation of such domains is predominantly determined and strongly pinned by the underlying crystal conditions, such as internal stress or disorder.

We now comment on the relevance of the present observation to other experimental results.  The nuclear magnetic resonance (NMR) spectra have been reported to exhibit an anomalous broadening in the hidden-order phase for {\boldmath $H$}$\parallel ab$ \cite{Mat01,Tak07,Ber01}.  This phenomenon has been discussed in terms of the orbital currents associated with the breaking of time-reversal symmetry \cite{Cha02}. The present results provide an alternative explanation to this: Finite values of $\chi_{ab}$ give rise to the broadening of the NMR spectrum, which is estimated to be $\sim$ 1\,Oe at 4\,T \cite{SOM}, roughly the same as the reported values \cite{Tak07}. We also note that some of the nuclear quadrupole frequency $\nu_Q$ measurements \cite{Ber06}, which are sensitive to the local charge density, report a small change in the slope of $\nu_Q(T)$ at the Ru site below $T_h$. As for the crystal structure, no discernible lattice distortions have been reported. This may be a result of the weak coupling between the lattice and electronic excitations relevant to the two-fold symmetry. The symmetry of superconductivity, which appears at $T_c=1.4$\,K and is embedded in the hidden-order phase, should also be restricted by the found rotational asymmetry. Recently, a chiral $d$-wave superconducting state of the form 
\begin{equation}
\sin\frac{k_z}{2}c\left(\sin\frac{k_x+k_y}{2}a+i\sin\frac{k_x-k_y}{2}a\right)
\end{equation}
has been proposed \cite{Kas07}. The rotational symmetry breaking implies that the states including $\sin\frac{k_x+k_y}{2}a$ and $\sin\frac{k_x-k_y}{2}a$ have different superconducting transition temperatures $T_{c1}$ and $T_{c2}(<T_{c1})$, which results in a phase transition inside the superconducting phase at $T_{c2}$. A recently reported anomaly in the lower critical field measurements appears to support such exotic superconducting states \cite{Oka10}.

Our results, together with the absence of a large magnetic ordered moment, impose constraints on theoretical models of the hidden order in URu$_2$Si$_2$. The breaking of the four-fold rotational symmetry in a tetragonal crystal structure is in general a hallmark of electronic nematic phases \cite{Kivelson98}. Electronic states that break crystal rotational symmetry without showing ordered moments have also been suggested in various strongly correlated electron systems, including Sr$_{3}$Ru$_{2}$O$_{7}$ \cite{Bor07}, high-$T_c$ cuprates \cite{Voj09}, and frustrated magnets \cite{Shannon06}, which are discussed in terms of stripe or nematic orders. Our finding of a directional electronic state in a heavy-fermion material is another example of such exotic order in correlated matter.


\end{document}